# QUALITY OF LIFE AND PERCEIVED CARE OF PATIENTS IN ADVANCED CHRONIC KIDNEY DISEASE CONSULTATIONS: A CROSS-SECTIONAL DESCRIPTIVE STUDY

CALIDAD DE VIDA Y DE LA ATENCIÓN PERCIBIDA DEL PACIENTE EN LA CONSULTA DE ENFERMEDAD RENAL CRÓNICA AVANZADA: ESTUDIO DESCRIPTIVO TRANSVERSAL

**AUTHORS:**

Verónica Gimeno-Hernán[1,2,3]

verogime@ucm.es  - ORCID: 0000-0002-7427-7833

María Isabel Duran-Muñoz[3]

iduran72@gmail.com

María Rosario Del Pino-Jurado[3]

mariarosariodel.pino@salud.madrid.org

Araceli Farlado-Cabana[1,2,4]

arafaral@ucm.es - ORCID: 0000-0001-9939-0034

Marta Oliva-Hernando[5]

ohmarta84@gmail.com

Ismael Ortuño-Soriano[1,2]

iortunos@ucm.es  - ORCID: 0000-0001-5651-8376

**AFFILIATIONS:**

1. Department of Nursing, Faculty of Nursing, Physiotherapy and Podology, Universidad Complutense de Madrid, Plaza Ramón y Cajal s/n, 28040 Madrid,
2. Spain. Health care research group, Instituto de Investigación Sanitaria Hospital Clínico San Carlos (Iddisc), 28040 Madrid, Spain.
3. Departament Haemodyalisis, Hospital Clínico San Carlos, 28040 Madrid, Spain.

4. Departament oncology. Hospital Puerta de Hierro, 28220, Majadahonda-Madrid, Spain.
5. Intensive Care Unit. Hospital de la Princesa, 28006, Madrid, Spain.

**CORRESPONDENCE AUTHOR:**

*Verónica Gimeno Hernán*

*C/ Profesor Martín Lagos SN, Planta baja-Hemodiálisis. 28040. Madrid. Madrid.*

*Movil Contact: +34 678 04 96 76 Email: verogime@ucm.es*

**WHAT IS KNOWN / WHAT IT CONTRIBUTES.**

This study highlights the importance of prioritizing quality of life and nursing care in renal patients. It investigates the relationship between patients' perceptions of care quality and their quality of life in an advanced chronic kidney disease (CKD) clinic. The results indicate that while both quality of life and perceived care quality are crucial, they are independent dimensions in these patients. It was found that factors like medication use and age influenced the perception of care quality, but no correlation was observed between quality of life and care quality. The study emphasizes the need for further research, including larger multicenter studies, to validate these findings and improve outcomes for advanced CKD patients.

# QUALITY OF LIFE AND PERCEIVED CARE OF PATIENTS IN ADVANCED CHRONIC KIDNEY DISEASE CONSULTATIONS: A CROSS-SECTIONAL DESCRIPTIVE STUDY

**ABSTRACT:**

Objetive: In the care of renal patients, prioritising their quality of life and nursing care is essential. Research links patients' perceptions of care quality to improved outcomes such as safety, clinical efficacy, treatment adherence, and preventive practices.

This study aimed to evaluate the quality of life and care perception in these patients and explore potential associations between these dimensions.

Material and methods: A cross-sectional descriptive study was conducted with 43 patients attending an advanced CKD clinic. Quality of life was assessed using the KDQOL-36 questionnaire, while the IECPAX questionnaire measured perceived care quality. Sociodemographic and clinical data were collected from patient records. Participants completed the questionnaires during routine visits, with scores analysed to identify associations between variables.

Results: The study included 60% men (n=28) and 32% women (n=15), with a mean age of 78 years (±16). Among participants, 45% were diabetic, 79% hypertensive, and 58% took more than five medications daily. Mean scores were 78.76 (±12.15) for KDQOL-36 and 5.54 (±2.64) for IECPAX. Significant differences were found in the physical role domain between men and women (p=0.01) and for individuals over 65 years (p=0.04). Higher IECPAX scores were associated with taking more than five medications (p=0.05). However, no correlation was observed between KDQOL-36 and IECPAX scores.

Conclusions: The findings suggest that quality of life and perceived care quality are independent in advanced CKD patients. While this study provides insights, larger multicentre studies are needed to validate these results. These findings highlight the importance of addressing both aspects separately to improve outcomes in this population.

# CALIDAD DE VIDA Y DE LA ATENCIÓN PERCIBIDA DEL PACIENTE EN LA CONSULTA DE ENFERMEDAD RENAL CRÓNICA AVANZADA: ESTUDIO DESCRIPTIVO TRANSVERSAL

**RESUMEN:**

Objetivo: Evaluar la calidad de vida y la percepción de la atención recibida en pacientes en consulta de enfermedad renal crónica avanzada.

Material y métodos: Se realizó un estudio descriptivo transversal con 43 pacientes con enfermedad renal crónica avanzada. La calidad de vida se evaluó mediante el cuestionario KDQOL-36, mientras que la calidad percibida de la atención se midió con el cuestionario IECPAX. Se recopilaron datos sociodemográficos y clínicos de los registros de los pacientes. Los participantes completaron los cuestionarios durante visitas rutinarias, y las puntuaciones fueron analizados para identificar asociaciones entre las variables.

RESULTADOS: El estudio incluyó un 60 % de hombres (n=28) y un 32 % de mujeres (n=15), con una edad media de 78 años (±16). Entre los participantes, el 45 % eran diabéticos, el 79 % hipertensos y el 58 % tomaban más de cinco medicamentos al día. Las puntuaciones promedio fueron de 78,76 (±12,15) para el KDQOL-36 y de 5,54 (±2,64) para el IECPAX. Se encontraron diferencias significativas en el dominio del rol físico entre hombres y mujeres (p=0,01) y en individuos mayores de 65 años (p=0,04). Las puntuaciones más altas de IECPAX se asociaron con tomar más de cinco medicamentos (p=0,05). Sin embargo, no se encontró relación significativa entre as puntuaciones de KDQOL-36 e IECPAX.

Conclusiones: Los hallazgos sugieren que la calidad de vida y la calidad percibida de la atención son independientes en pacientes con ERCA avanzada. Estos hallazgos resaltan la importancia de abordar ambos aspectos por separado para mejorar los resultados.

**Palabras clave: Calidad de vida, enfermería, cuidado y enfermedad renal.**

**INTRODUCTION:**

The concepts of quality of life and patient-centred care are relatively recent terms defined by the World Health Organization (WHO) [1,2]. Quality of life is understood as an individual's personal perception of their life situation within the cultural context in which they live and in relation to their expectations, values, and interests[1]. Patient-centred care, on the other hand, is a concept that goes beyond the traditional scope of care, defined as care that views the individual as a whole, encompassing various levels of needs and objectives that stem from the person's own health determinants[2].

Advances in medicine in recent years have led to increased longevity and a growing number of individuals living with chronic illnesses. As a result, there is an increasing need to focus on quality of life and the integrated care of patients. These concepts are now regarded as essential components in the evaluation of health outcomes, with the aim not only of achieving clinical stability but also of addressing patients' broader perceptions and needs[3].

Advanced chronic kidney disease (ACKD) represents the final stage of chronic kidney disease[4,5], characterised by a glomerular filtration rate of less than 15 ml/min due to structural or functional kidney damage. Patients with ACKD often stand on the threshold of initiating renal replacement therapy.

Chronic kidney disease (CKD) tends to become more prevalent with advancing age and is often accompanied by a range of comorbidities and complications. These factors significantly impact the social, familial, and occupational spheres of those affected, leading to a substantial decrease in their quality of life[6]. It has been observed that improved management of quality of life, supported by the efforts of multidisciplinary teams in specialised units, can enhance patients' ability to cope with the disease during this critical stage[7].

Furthermore, the quality of care provided to patients has emerged as a pivotal factor. Evidence suggests that patient experience is positively associated with safety and clinical effectiveness across a wide range of diseases, care settings, population groups, and outcome measures[8]. Additionally, positive associations have been found between a good care experience and adherence to recommended treatment and preventive care practices—both crucial aspects for renal patients[8].

This context underscores the need for a study to assess the quality of life and the quality of care provided to patients in ACKD consultations. Such an evaluation would allow for the identification of aspects that influence patients' daily functioning and their perception of the care received.

## OBJECTIVES

The primary aim of the study was to evaluate the quality of life and the perception of medical care received by patients with advanced chronic kidney disease (ACKD) in a hospital setting. Through this evaluation, the study sought to identify potential associations between these two dimensions and their relationship with the patients' sociodemographic and health variables.

Specifically, the secondary objectives included: describing the results obtained in the various dimensions of the **KDQOL-36** questionnaire, designed to measure health-related quality of life in kidney disease patients[9,10,11,12]; analysing the dimensions of the **IECPAX** questionnaire, which measures the experience of perceived care[13]; and examining the associations between variables such as gender, age, comorbidities, and the outcomes of both instruments.

## MATERIALS AND METHODS

**Study design**

A cross-sectional descriptive study was conducted, including patients with ACKD who attended consultations at a tertiary hospital between January and July 2023. This approach provided a precise snapshot of patients' conditions during the study period.

**Population and selection criteria**

Patients aged 18 years or older, diagnosed with stage V ACKD, and attending specific consultations for this condition were included. Inclusion criteria required patients to have a glomerular filtration rate below 15 ml/min and be on the verge of initiating renal replacement therapy. Only those who understood and signed the informed consent form were included. Patients in earlier stages of the disease or those who commenced renal replacement therapy during the study period were excluded.

**Evaluation instruments**

Patients completed two questionnaires:

1. **KDQOL-36**[9,10,11,12]: This instrument, adapted into Spanish, consists of 36 questions measuring quality of life across various health-related dimensions. Scores range from 0 to 100, with higher scores indicating better health status.

2. **IECPAX**[13]: Developed in Spain, this questionnaire evaluates the experience of chronic patients with medical care, focusing on aspects such as productive interactions, self-management, and new relationships with the healthcare system.

Both questionnaires were self-administered, though assistance from the consultation nurse was available for patients with comprehension or reading difficulties.

**Data collection process**

The purpose of the study was explained to patients, who were provided with the questionnaires, an information sheet, and an informed consent form. Participants completed the questionnaires during their regular ACKD consultation visits. Concurrently, sociodemographic and health data were collected from clinical records, including information on comorbidities such as diabetes and hypertension.

**Data analysis**

Data were analysed using **SPSS 26.0**. Quantitative variables were summarised using measures of central tendency and dispersion, while qualitative variables were presented as percentages. Statistical tests such as the Student's *t* test, ANOVA, or Mann-Whitney test were applied depending on data normality. Associations were considered significant at a 95% confidence level.

**Ethical considerations**

The study adhered to the principles of the Declaration of Helsinki and Good Clinical Practice guidelines, ensuring data confidentiality and participant anonymity. Ethical approval was obtained from the relevant ethics committee (CI-23038-E), and The participants voluntarily participated by signing the informed consent form.

**RESULTS:**

A total of 43 patients who adequately completed the questionnaires were included. Of these, 60% (n=28) were men and 32% (n=15) were women, with an average age of 78 years (±16). Among them, 45% (n=21) were diabetic, and 79% (n=37) had hypertension. A total of 58% (n=25) reported taking more than five medications daily. The mean scores obtained from both questionnaires were 78.76 (±12.15) on the KDQOL-36 and 5.54 (±2.64) on the IEXPAC.

The data regarding the distribution of scores on both questionnaires are presented in Figures 1 and 2.

Results of the KDQOL-36 Questionnaire:
In this study, 42 individuals were evaluated, and their average scores were analysed in relation to various characteristics, such as sex, age, and the presence of certain conditions. The p-value was used to determine whether the differences observed between groups were significant, as shown in the results in Table 1.

Hypertension was the only factor that showed a significant difference in patients' quality of life. Patients without hypertension reported a better quality of life than those with hypertension. Other factors, such as sex, age, diabetes, cardiac issues, and the number of medications, did not show significant differences in this study.

In most dimensions of the KDQOL-36 questionnaire, no significant differences were found between men and women. The only exception was the "Emotional Role" dimension, where women reported a better quality of life than men, with a statistically significant difference (p-value = 0.03) (Table 2). This suggests that, emotionally, women with kidney disease may feel less affected than men. In other dimensions (physical, physical role, bodily pain, general health, vitality, social function, and mental health), the differences observed were not significant, indicating that the health-related quality of life for men and women with kidney disease is quite similar according to this study.

Overall, the results in Table 3 show that age does not have a significant impact on most dimensions of the KDQOL-36 questionnaire among patients with chronic kidney disease. However, a significant difference was observed in the vitality dimension, where patients under 65 reported higher levels compared to older patients. This suggests that the perception of energy and vitality may vary with age in this specific population.

Results of the IEXPAC Questionnaire:

Table 4 presents the results of the IEXPAC questionnaire, which measures a specific aspect of patients' quality of life, showing mean scores and standard deviations for different subgroups within a sample of 42 individuals. Factors such as sex, age, diabetes mellitus, hypertension, cardiovascular pathology, and the number of medications were examined.

The data from the IEXPAC questionnaire indicate significant differences concerning sex, with men achieving higher scores. There were no significant differences among the other subgroups analysed for age, presence of diabetes mellitus, hypertension, cardiovascular pathology, and the number of medications, suggesting that these factors do not significantly influence IEXPAC scores in this sample of 42 individuals.

Regarding sex, the results in Table 5 indicate that it does not significantly influence most dimensions of the IEXPAC questionnaire, except for the "New Relational Model" dimension, where men perceived better adaptation compared to women.

The data in Table 6 suggest that age does not seem to significantly influence the perception of care received in the advanced chronic kidney disease nursing consultation, as measured by the IEXPAC questionnaire dimensions.

Concerning the main objective of the study, the data obtained indicate that the scores of the KDQOL-36 and IEXPAC questionnaires have distinct ranges and averages, with the KDQOL-36 focusing on health-related quality of life and the IEXPAC on the experience of care received. No statistically significant correlation (p-value = 0.46) was found between the scores of the two questionnaires in this sample of 42 patients (Table 7).

**DISCUSSION:**

From the earliest studies on health-related quality of life (HRQoL) in our country to the present, renal replacement therapies (haemodialysis or peritoneal dialysis) remain the most frequently examined variables in HRQoL research for advanced chronic kidney disease (ACKD) patients. Most studies show greater impairment in patients undergoing replacement therapy compared to those in pre-dialysis stages[14,15,16]. Additionally, numerous studies highlight a strong association between symptoms of anxiety and depression and poorer HRQoL perceptions, which manifest to varying degrees in all ACKD patients[14,15,16,17,18,19].

The comorbidities associated with ACKD are also consistently identified as significant determinants of HRQoL[19,20]. In this study, hypertension emerged as a significant factor influencing patients' quality of life, aligning with existing literature[21]. For example, Lo et al. reported poorer quality of life among CKD patients with hypertension due to the additional burden of complications[22].

This study also found that women reported better HRQoL in the "Emotional Role" dimension than men, corroborating findings by Butt et al., who observed that women often exhibit better emotional resilience in chronic illnesses. However, the lack of significance in most other KDQOL-36 dimensions, based on factors like gender, age, and medication count, contrasts with other studies. Mujais et al., for instance, found that polypharmacy negatively affects HRQoL in CKD patients due to side effects and the complexities of managing multiple medications[23].

Perceived quality of care, as measured by the IECPAX questionnaire, showed that men reported better care experiences, particularly in the "New Relational Model" dimension. This aligns with findings by Hughes et al., who noted that men tend to report better nursing care experiences due to differing expectations and communication styles[24]. However, these findings contrast with Rodríguez et al., who found that women often report better care experiences due to a greater likelihood of active engagement in their care[25].

However, other similar studies conducted along the same lines, using the IEXPAC questionnaire in patients with rheumatic diseases (5.5 points) or type II diabetes mellitus (5.9 points), show results very similar to those found in our study, possibly due to similar sociodemographic characteristics[27,28].

On the other hand, the results related to the dimensions of the "New Relational Model" received the lowest scores, consistent with those reported by Fernández-Díaz et al., potentially due to the average age of the patients attending these consultations[26].

Items related to patient self-management yielded positive results. Nuño et al. highlight self-care in chronic patients as a factor of empowerment[29], which is possibly linked to the continuous follow-up by the same nurse in the ACKD consultation, as demonstrated by the findings of Fernández-Díaz et al[26].

The lack of a significant correlation between KDQOL-36 and IEXPAC scores suggests that these instruments measure different aspects of the patient experience. This conclusion aligns with other studies that have also reported low correlation between measures of quality of life and satisfaction with medical care. Specifically, a study by Feroze et al. emphasised that the perception of care received may not be directly related to perceived quality of life, as different factors influence these dimensions[30].

**Study limitations:**

From a methodological standpoint, one of the main limitations is the sample size. Despite efforts to include the entire possible population of patients, it would be desirable to increase the number of patients studied. This could only be achieved through a similar multicentre study or by extending the study period in the same unit.

**Sources of funding:**

This research has not received specific funding from public sector agencies, commercial sector, or non-profit organizations.

## CONCLUSIONS:

This study provides valuable insights into the quality of life and perceived care in patients with advanced CKD. Hypertension stands out as a significant factor affecting quality of life, while gender and age have minimal impact on most dimensions evaluated. Perceived care experience varies by gender, particularly in the "New Relational Model" dimension.

Further research is essential to better understand the factors influencing HRQoL and perceived care in this population, aiming to improve interventions and care delivery. These findings can guide healthcare professionals in implementing tailored strategies that address the specific needs of CKD patients, thereby improving both their quality of life and satisfaction with care.

**FIGURES:**

*Figure 1. Distribution of scores obtained in the KDQOL-36 questionnaire.*

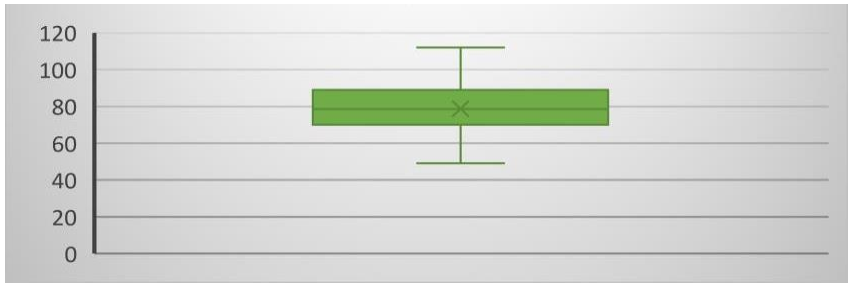

*Figure 2. Distribution of scores obtained in the IEXPAC questionnaire.*

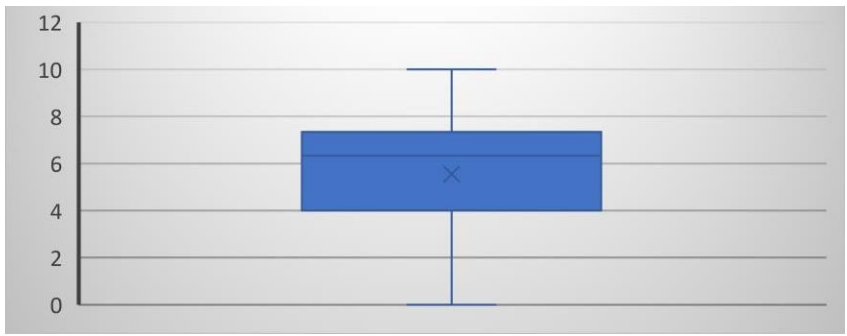

**TABLES:**

1: Analysis of the influence of sociodemographic and comorbidity variables on KDQOL-36 scores.

| KDQOL-36 Questionnaire Scores | | | |
|---|---|---|---|
| | | N=42 Mean, standard Desviation | p-value |
| Sex | Male | 77,6 (11,7) | 0,42 |
| | Female | 80,8 (12,9) | |
| Age | >65 years | 77,74 (11,7) | 0,22 |
| | <65 years | 83,9 (15,7) | |
| Diabetes mellitus | NO | 82,1 (10,6) | **0,06** |
| | YES | 75,1 (13) | |
| Hypertension arterial | NO | 92 (12,3) | **0,00** |
| | YES | 76,6 (10,7) | |
| Cardiovascular disease | NO | 78,4 (8,7) | 0,90 |
| | YES | 78,9 (14,7) | |
| Number of medications | >6 | 80 (10,6) | 0,29 |
| | <6 | 75 (8) | |
| Number of medications | >5 | 79,62 (10,3) | 0,38 |
| | <5 | 74,7 (9,7) | |

*2: Influence of gender on scores across KDQOL-36 dimensions.*

| KDQOL- 36 | Male n=28 Mean, St. Desviation | Female n=15 Mean, St. Desviation | p-value |
|---|---|---|---|
| **Physical Dimension (10 items)** | 20,7 (5,3) | 21,1 (2,9) | 0,79 |
| **Physical Role (4 items)** | 13,3 (4,6) | 12,8 (1,7) | 0,71 |
| **Bodily Pain (2 items)** | 6,5 (2,8) | 5,3 (2,2) | 0,16 |
| **General Health (5 items)** | 9,6 (4,1) | 11,3 (4) | 0,19 |
| **Vitality (4 items)** | 8,4 (3,2) | 9,9 (3,9) | 0,18 |
| **Social Function ( 2 items)** | 1,46 (1,0) | 1,7 (1,1) | 0,55 |
| **Emotional Role (3 items)** | 5,4 (2,7) | 7,4 (2,9) | **0,03** |
| **Mental Health (5 items)** | 9,54 (4,7) | 11,2 (4,7) | 0,26 |

*3: Influence of age on KDQOL-36 scores for advanced chronic kidney disease patients.*

| Items KDQOL-36 Questionnaire | | | |
|---|---|---|---|
| | Age | | p-value |
| | >65 years | <65 years | |
| | Mean, standart desviation | | |
| **Physical Dimension (10 items)** | 20,6 (5) | 22 (1,2) | 0,46 |
| **Physical Role (4 items)** | 13,3 (4,1) | 12,1 (1,2) | 0,45 |
| **Bodily Pain (2 items)** | 6,3 (2,8) | 5,1 (2) | 0,3 |
| **General Health (5 items)** | 9,8 (3,9) | 12,1 (5,1) | 0,18 |
| **Vitality (4 items)** | 8,4 (3) | 11,3 (4,8) | **0,04** |
| **Social Function ( 2 items)** | 1,4 (0,9) | 2,1 (1,6) | 0,09 |
| **Emotional Role (3 items)** | 5,8 (2,7) | 7,4 (3,8) | 0,18 |
| **Mental Health (5 items)** | 9,9 (4,4) | 11,6 (6,3) | 0,51 |

*4: Influence of gender on IECPAX scores for advanced chronic kidney disease patients.*

|  |  | IECPAX Questionnaire score | |
|---|---|---|---|
|  |  | N=42 | p-value |
|  |  | Mean, st. desviation | |
| **Sex** | Male | 5,74 (2,9) | **0,05** |
|  | Female | 5,18 (1,9) | |
| **Age** | >65 years | 5,6 (2,5) | 0,50 |
|  | <65 years | 4,9 (3,4) | |
| **Age** | >75 years | 5,8 (2,7) | 0,41 |
|  | <75 years | 5,1 (2,6) | |
| **Diabetes mellitus** | NO | 5,7 (2,9) | 0,77 |
|  | YES | 5,4 (2,3) | |
| **Hypertension arterial** | NO | 4,2 (3,6) | 0,20 |
|  | YES | 5,7 (2,5) | |
| **Cardiovascular disease** | NO | 5,9 (2,5) | 0,47 |
|  | SI | 5,3 (2,7) | |
| **Number of medications** | >6 | 6,7 (1,5) | 0,51 |
|  | <6 | 6,2 (2,5) | |
| **Number of medications** | >5 | *6,6 (1,4)* | *0,96* |
|  | <5 | *6,5 (3,1)* | |

5: Analysis of the influence of sociodemographic and comorbidity variables on IECPAX scores.

| **IEXPAC Questionnaire Items** | | | |
|---|---|---|---|
| **Dimensions** | **Male n=28** | **Female n=15** | **p-value** |
| | Mean, standart desviations | | |
| **Productive interactions (1,2,5 and 9)** | 7,6 (3,7) | 8,1 (2,9) | 0,68 |
| **New relational model ( 3,7 and 11)** | 5,6 (3,4) | 3,3 (2,8) | **0,03** |
| **Patient self-management (4,6,8 and 10)** | 7,2 (3,6) | 7,4 (2,7) | 0,83 |

6: Influence of age on IECPAX scores for advanced chronic kidney disease patients.

| **Items Questionnaire IEXPAC** | | | |
|---|---|---|---|
| | **AGE** | | **p-value** |
| | **>65 years** | **<65 years** | |
| | Mean, standart desviation | | |
| **Productive interactions (1,2,5 and 9)** | 7,8 (3,1) | 7,1 (4,9) | 0,61 |
| **New relational model ( 3,7 and 11)** | 4,8 (3,3) | 5,0 (3,8) | 0,86 |
| **Patient self-management (4,6,8 and 10)** | 7,4 (3,1) | 6,7 (4,6) | 0,59 |

7: Relationship between KDQOL-36 and IECPAX scores in advanced chronic kidney disease nursing consultations.

|  | Measure | |
|---|---|---|
|  | Mean, standart desviation | p-value |
| **KDQOL-36 Questionnaire Score (N=42)** | 78,7 (12,1) | **0,46** |
| **IEXPAC Questionnaire Score (N=42)** | 5,48 (2,6) | |